\documentclass[usenatbib]{mn2e}

\usepackage{graphicx}
\input{epsf}
\usepackage{epsf}
\usepackage{dcolumn}

\newcommand{\be}{\begin{equation}}
\newcommand{\ee}{\end{equation}}
\newcommand{\bea}{\begin{eqnarray}}
\newcommand{\eea}{\end{eqnarray}}

\begin{document}
\title[]
{Evolution of Spherical Overdensity in Thawing Dark Energy Models}
\author[]
{N.~Chandrachani Devi$^1$\thanks{email:chandrachani@gmail.com }
 and Anjan A Sen,$^1$\thanks{email:anjan.ctp@jmi.ac.in} \\
$^1$Center For Theoretical Physics,
Jamia Millia Islamia, New Delhi 110025, India}

\date{\today}
\pagerange{\pageref{firstpage}--\pageref{lastpage}} \pubyear{2010}

\maketitle

\begin{abstract}
We study the general evolution of spherical over-densities for thawing class of dark energy models. We model dark energy with scalar fields having  canonical as well as non-canonical kinetic energy. For non-canonical case, we consider models where the kinetic energy is of the Born-Infeld Form. We study various potentials like linear, inverse-square, exponential as well as PNGB-type. We also consider the case when dark energy is homogeneous as well as the case when it is inhomogeneous and virializes together with matter. Our study shows that models with linear potential in particular with Born-Infeld type kinetic term can have significant deviation from the $\Lambda$CDM model in terms of density contrast at the time of virialization. Although our approach is a simplified one  to study the nonlinear evolution of matter overdensities inside the cluster and is not applicable to actual physical situation, it gives some interesting insights into the nonlinear clustering of matter in the presence of thawing class of dark energy models.

\end{abstract}


\section{Introduction}
One of the most significant discoveries in cosmology in recent years is the fact that our universe is currently going through an accelerated expansion phase\citep{Riess2004,Knop2003}.
This can have far reaching
implications for fundamental theories of physics. This late time acceleration of the universe can be due to the presence of an exotic fluid with
large negative pressure known as {\it dark energy} or due to the modification of
gravity itself. The simplest candidate of dark energy is provided by
cosmological constant $\Lambda$ with equation of state parameter $w=-1$.
However, the $\Lambda$CDM model  is plagued with
fine tuning and cosmic coincidence problems (See \cite{Copeland2006,Sami2009,Sahni2000,Padmanabhan2003,Linder2008,Frieman2008,Caldwell2009,Silvestri2009}
 for a nice review).

Scalar field models mimicking a variable $\Lambda$ can alleviate the fine
tuning and coincidence problems and provide an interesting
alternative to cosmological constant\citep{Ratra1988,Caldwell1998,Liddle1999,Steinhardt1999}. These 
 scalar field models can be  broadly classified
into two categories depending upon the form of their potentials:
 fast roll and slow roll models termed as freezing
and thawing models in the literature\citep{Caldwell2005}.  In case of 
fast roll models,
the potential is steep resulting the scalar field to track  the
background fluid and is sub dominant for most of the evolution history. Only
at late times, the field becomes dominant and drives the
acceleration of the universe. Such solutions are also known as {\it
trackers}.

Then there are the slow-roll models for which the field kinetic energy is
much smaller than its potential energy. Usually it has 
sufficiently flat potential similar to an inflaton. At early times,
the field is nearly frozen at $w=-1$ due to the large Hubble friction.
Its energy density is nearly constant with a negligible
contribution to the total energy density of the universe. But as 
radiation/matter rapidly dilutes due to
the expansion of the universe and the background energy density
becomes comparable to scalar field energy density, the field breaks away
from its frozen state and evolves slowly to the region with $w > -1$. 
However, in this case, the model needs some degree of fine tuning of the initial conditions in order to achieve a viable late time evolution.

Recent observations suggest that the equation of state parameter for
dark energy does not significantly deviate from $w=-1$ around the present epoch
\citep{Wood-Vasey2007,Davis2007}.
This type of equation of state can be easily obtained in dynamical
models represented by thawing scalar fields. Motivated by this fact, 
Scherrer and Sen \citep{Scherrer2008a} examined quintessence models
with nearly flat potentials satisfying the slow-roll conditions. It was
shown that under the slow-roll conditions,  a scalar field with a variety
of potentials $V(\phi)$ evolves in a similar fashion and one can derive a
generic expression for the equation of state for all
  such scalar fields. Similar results were later established for the 
case of phantom \citep{Scherrer2008b}
and tachyon scalar fields \citep{Amna2009,soma2010}. It was demonstrated that under slow-roll conditions, all of them have identical equation of state and hence can
not be distinguished, atleast, at the level of background cosmology. The crucial assumption for
arriving  at this important conclusion was the fulfillment of the slow-roll conditions for the field potentials.

In a recent paper by Sen et al. \citep{soma2010}, assumption of slow roll has been relaxed, 
but it was assumed that
the scalar field is of thawing type i.e it is initially frozen at
$w=-1$ due to large Hubble damping. With these choices the evolution
 of a variety of scalar field models having both canonical and non-canonical kinetic terms have been studied. Observational quantities like Hubble parameter, luminosity
 distance and quantities related to the Baryon Acoustic Oscillation (BAO) measurement have been studied for variety of potentials. It was shown that given the current error bars for different observational data related with background cosmology, it is difficult to distinguish different thawing models from $\Lambda$CDM. It was also shown that lower the value of $\Omega_{m0}$ parameter, higher the chance to distinguish and there is a redshift range between $z \sim 0.4 - 1$ where the deviation from $\Lambda$CDM is maximum irrespective of the model considered. 

Dark energy not only affects background expansion rate and thereby modifying the observables like distance-redshift relation, it also affects the growth of structure. On large scales, dark energy also clusters which in turn affects the matter clustering. But at small scales, although dark energy is more or less smoothly distributed, it provides an extra Hubble drag thereby slowing down the growth rate of matter perturbations. The behavior of linear perturbation in a scalar field  and its effect on large scale structure of the universe have been studied by a  number of authors (\citet{Bartolo2004,Hu2005,Gordon2005,Avelino2008,Unnikrishnan2008a,Gordon2004,Zimdahl2005,Unnikrishnan2008b,Jassal2009,Jassal2010}). The impact has been studied for cosmic microwave background \citep{Macorra2003,Baccigalupi2006,Acquaviva 2006,Giovi2005}, galaxy redshift surveys (\citet{Eisenstein1998,Seo2003,Haiman2000,Jain2003,Bernstein2004,Allen2004}), cross-correlation of the integrated Sachs-Wolf effect \citep{Pogosian2005,Corasaniti2005,Giannantonio2008}as well as in neutral hydrogen surveys \citep{Bharadwaj2009}.

In the non-linear regime one has to perform N-Body simulations to accurately measure the effect of dark energy on the clustering properties of matter and this has been investigated in a number of papers \citep{Maccio2004,Baldi2008,Couetin2010}. The simplest analytical approach to study the nonlinear  clustering of matter is the spherical collapse formalism first developed by Gunn and Gott \citep{Gunn1972}. It describes how a small spherical
over density decouples from the background evolution, slows down, then 
eventually turns around and collapses. It is generally assumed that
the collapse is not complete so the system does not reach the
singularity, instead it eventually virializes and stabilizes to a
finite size. Combining this with the Press-Schechter formalism \citep{Press1974}, one can have a model for formation of structures which predicts
 the abundances of virialized  objects as a function of mass. This 
model can give insight into the physics of structure formation which can be
ultimately used for a detailed N-Body simulation.  There are numerous studies of the spherical collapse formalism to include both homogeneous and
inhomogeneous dark energy \citep{Lahav1991,Shaw2008,Manera2006,Lokas2001,Basilakos2007,Schaefer2008,Bartelmann2006,Basilakos2003,Horellou2005,Maor2005,Mota2004,Nunes2006,Wang1998,Wang2006,Weinberg2003,Zeng2005}.

In this work, we systematically study the spherical collapse model for thawing type scalar field dark energy models for various potentials. We consider the canonical as well as noncanonical kinetic energy for the scalar fields. We assume the dark energy to be  homogeneous as well as inhomogeneous when it also virializes together with matter inside the spherical over density. We calculate the matter density contrast at turnaround as well as at virialization and compare the result with the corresponding $\Lambda$CDM model. 

The plan of the paper is as follows: in section 2 we describe the background evolution of thawing dark energy models for ordinary scalar field as well as for tachyons. In section 3 we study the evolution of linear and nonlinear matter density perturbation assuming the dark energy to be homogeneous. In section 4 we describe the spherical collapse formalism for a general scalar field dark energy model and study the virialization process in this scenario and discuss the results obtained. Finally, we draw our conclusions in section 5.  
\vspace{5mm}
\section{Background Evolution for Thawing Dark Energy Models}
As discussed in the introduction, thawing dark energy models are characterizedin such a way that in the early universe the scalar field is frozen by  very large Hubble damping due to the expansion of the universe. As the universe expands the Hubble parameter decreases so the Hubble damping and the scalar field starts evolving slowly down its potential. Therefore, the equation of state initially starts with $w=-1$ and slowly departs from this value in the later time.
In this paper we consider both ordinary scalar field with canonical kinetic term as well as tachyon type scalar field having Born-Infeld type kinetic term.
\vspace{2mm}
\subsection{Thawing Scalar field}
 In what follows, we shall assume that the dark energy is described
by a minimally-coupled scalar field, $\phi$, with equation of motion
\begin{equation}
\ddot{\phi}+3 H{\phi}+ dV/d\phi = 0 \label{motionq}
\end{equation}
where the Hubble parameter $H$ is given by
\begin{equation}
\label{H}
H = \left(\frac{\dot{a}}{a}\right) = \sqrt{ 8 \pi G\rho/3}.
\end{equation}
Here $\rho$ is the total energy density in the universe. We model a
flat universe containing only matter and a scalar field, so that
$\Omega_\phi + \Omega_M = 1$.

Equation (\ref{motionq}) indicates that the field rolls downhill in
the potential $V(\phi)$ but its motion is damped by a term
proportional to $H$. The equation of state parameter $w$ is given by
$w=p_{\phi}/\rho_{\phi}$ where the pressure and density of the
scalar field have the forms
\begin{eqnarray}
&&p_{\phi} = \frac{\dot \phi^2}{2} - V(\phi) ,\\
&&\rho_{\phi} = \frac{\dot \phi^2}{2} + V(\phi)
\end{eqnarray}

Defining $\lambda=-\frac{1}{V}\frac{dV}{d\phi}$ and $\Gamma \equiv V \frac{d^2 V}{d\phi^2}/\left(\frac{dV}{d\phi} \right)^2$, one can form an autonomous system
of equations involving the two observables $\Omega_{\phi}$ and $\gamma = (1+w)$ together with the parameter $\lambda$. This is derived by  \citep{Scherrer2008a}as
\begin{eqnarray}
\label{gammaprime}
\gamma^\prime &=& -3\gamma(2-\gamma) + \lambda(2-\gamma)\sqrt{3 \gamma
\Omega_\phi},\\
\label{Omegaprime}
\Omega_\phi^\prime &=& 3(1-\gamma)\Omega_\phi(1-\Omega_\phi),\\
\label{lambda}
\lambda^\prime &=& - \sqrt{3}\lambda^2(\Gamma-1)\sqrt{\gamma \Omega_\phi}.
\end{eqnarray}

Given the initial conditions for $\gamma$, $\Omega_{\phi}$ and $\lambda$, one can solve this system of
equations numerically for different potentials. We are interested in thawing models
i.e models for which the equation of state is initially frozen at $w=-1$. Hence, we take $\gamma = 0$ initially
for our purpose. Also one has to choose the initial value $\lambda_{i}$. It was earlier shown that if $\lambda_{i}$ is small (which is equivalent to assuming the slow-roll conditions similar to inflation) the scalar field evolution is similar to that of cosmological constant $\Lambda$ \citep{Scherrer2008a}. Here we assume that the slow-roll conditions are strongly broken i.e $\lambda_{i} = 1$ in order to have a behavior which is in principle different from $\Lambda$.
In general the contribution
  of the scalar field to the total energy density of the universe is insignificant at early times, nevertheless
  one has to fine-tune the initial value of $\Omega_{\phi}$ in order to have its correct contribution at present.
  This is the fine tuning one needs to have in a thawing model. With these initial conditions we evolve the above
  system of equations from redshift $z=1000$ (or $a = 10^{-3}$) till the present day $z=0$ ($a=1$).
We consider various types of potentials e.g $V=\phi$, ~$V = \phi^2$,
~ $V=e^\phi$ and $V = \phi^{-2}$, characterized by
$\Gamma=0,~\frac{1}{2},~1$ and $\frac{3}{2}$ respectively.
 We assume 
$\Omega_{\phi0}=0.75$ at the present epoch for all chosen values of
$\Gamma$.

We also consider the Pseudo-Nambu Goldstone Boson (PNGB) model \citep{Frieman1995}.  (For a recent
discussion, see Ref. \citep{Abrahamse2007} and references therein).
This model is characterized by the potential
\begin{equation}
V(\phi) = m^4 [\cos(\phi/f)+1].
\end{equation}
Alam et al., \citep{Alam2003} have previously considered such type of potential to see whether dark energy is decaying or not.

Without any lose of generality, we choose $f=1$. Here, the constant $m$ is related to the mass of the field.
\vspace{2mm}
\subsection{Thawing Tachyon Model}
In recent years, tachyon-like scalar field having kinetic energy of Born-Infeld form has generated lot of interests in cosmology. There have been several investigations using this type of field as dark energy \citep{Copeland2005,Tsujikawa2004,Sami2004,Felder2002,Abramo2003,Bagla2003,Aguirregabiria2004}. In what follows, we form the similar set of equations for tachyon field as described above for ordinary scalar field.

The tachyon field is specified by the Dirac-Born-Infeld (DBI) type
of action\citep{Sen2002a,Sen2002b,Garousi2000,Bergshoeff2000,Kluson2000,Kutasov2003}:

\begin{equation}
{\mathcal{S}}=\int {-V(\phi)\sqrt{1-\partial^\mu\phi\partial_\mu\phi}}\sqrt{-g} d^4x.
\label{Taction1}
\end{equation}
In FRW background, the pressure and energy density of the tachyon field $\phi$
are given by
\begin{equation}
p_{\phi}=-V(\phi)\sqrt{1-\dot{\phi}^2}
\end{equation}
\begin{equation}
\rho_{\phi}=\frac{V(\phi)}{\sqrt{1-\dot{\phi}^2}}
\end{equation}
The equation of motion which follows from (\ref{Taction1}) is
\begin{equation}
\ddot{\phi}+3H\dot{\phi}(1-\dot{\phi}^2)+\frac{V'}{V}(1-\dot{\phi}^2)=0
\end{equation}
where $H$ is the Hubble parameter and here prime denotes the derivative with respect to $\phi$ . Now assuming 
$\lambda=-\frac{\frac{dV}{d\phi}}{V^{\frac{3}{2}}}$ and $\Gamma=V\frac{\frac{d^{2}V}{d\phi^2}}{(\frac{dV}{d\phi})^2}$, we can also form an autonomous system of equations involving $\Omega_{\phi}$, $\gamma = (1+w)$ and $\lambda$ for tachyon \citep{Amna2009}:

\begin{eqnarray}
\gamma'=-6\gamma(1-\gamma)+2\sqrt{3\gamma\Omega_{\phi}}\lambda(1-\gamma)^\frac{5}{4}
\label{tachgamma}\\
\Omega_{\phi}'=3\Omega_{\phi}(1-\gamma)(1-\Omega_{\phi})
\label{tachomega}\\
\lambda'=-\sqrt{3\gamma\Omega_{\phi}}\lambda^2(1-\gamma)^\frac{1}{4}(\Gamma-\frac{3}{2})
\label{tachlambda}
\end{eqnarray}

\begin{figure}
\includegraphics[width=80mm]{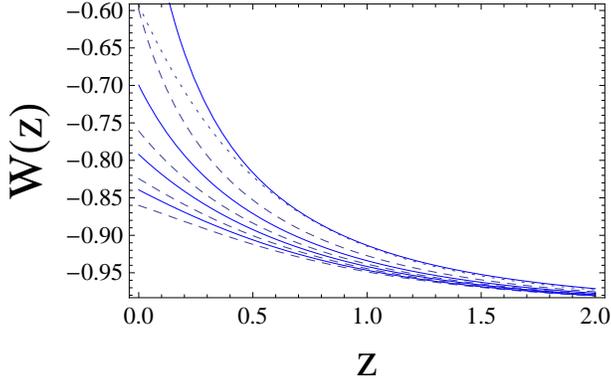}
\caption{Plot of equation of state $w$ vs. redshift for different scalar field
and tachyon models.
Solid curves represent different Tachyon models with $V(\phi) = \phi, \phi^2, e^{\phi}, \phi^{-2}$ respectively from top to bottom,
Dashed curves from top to bottom represent different scalar field models with same
potentials as in tachyon. Dotted curve represents PNGB model. $\Omega_{m0} = 0.25$.}
\end{figure}

We choose the similar set of initial conditions to solve this system of equations for tachyon as we describe earlier for ordinary scalar field. We also assume the same set of potentials $V=\phi$,~$V = \phi^2$,~
$V=e^\phi$ and $V = \phi^{-2}$, characterized by
$\Gamma=0,~\frac{1}{2},~1$ and $\frac{3}{2}$ respectively and $\Omega_{\phi0} = 0.75$.
\vspace{2mm}
\subsection{Background Result}
Let us now see the behavior of the different dark energy models that we have considered above. In Figure 1, we plot the behaviors of the equation of state parameter $w$ for different thawing models. 
It shows that the equation of state of different fields
with different potentials behave differently as one approaches the
present day although in the past their behaviors are almost identical.
This is not surprising as we have assumed the violation of
slow-roll condition, i.e $\lambda_{initial} \sim 1$. With slow-roll
condition satisfied, i.e, $\lambda_{i} << 1$, it was shown earlier
that models with different potentials have the identical $w(a)$ both
for scalar and tachyon fields\citep{Scherrer2008a,Scherrer2008b,Amna2009}. Although we show equations of state for some models which are more than $w= -0.8$ at present, these models are practically ruled out by current observational data.

\section{Evolution of Density perturbations}

\begin{figure}
\includegraphics[width=80mm]{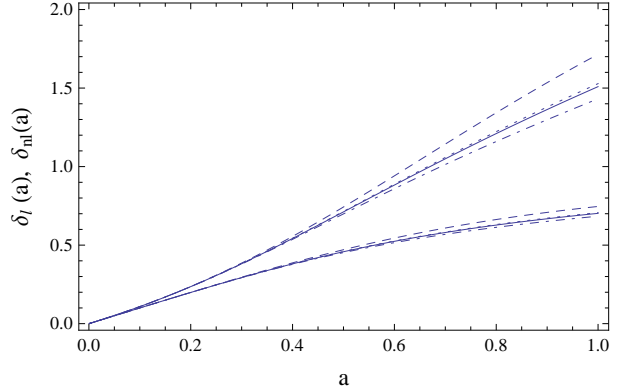}
\caption{Variation of linear and nonlinear density contrast $\delta$ as a function
  of scale factor for scalar field
and tachyon models.
Dashed line is for $\Lambda$CDM model, solid line is for scalar field with linear potential and dash-dotted line is tachyon field with linear potential. Dotted line is for scalar field with PNGB type potential. The upper set is the solution for full nonlinear equation given by eqn (16), where as the lower set is for linearized equation (17). $\Omega_{m0} = 0.25$.}
\end{figure}

The matter density contrast  is defined by $\delta = (\rho_{mc} - \rho_{m})/\rho_{m}$, where $\rho_{mc}$ is the perturbed matter energy density and $\rho_{m}$ is the background matter energy density. Assuming that the dark energy does not cluster, the evolution of the matter density contrast is governed by the equation
\begin{equation}
 {\ddot{\delta}}  +  2{\dot{a}\over{a}}{\dot{\delta}} -  
4\pi G \rho_{m}(1+\delta)\delta-{4\over{3}}{{\dot{\delta}}^2\over{(1+\delta)}}  = 0.
\end{equation}
In the linear regime, one can ignore higher order term in the above equation and approximate it as:
\begin{equation}
 {\ddot{\delta}}  +  2{\dot{a}\over{a}}{\dot{\delta}} -  
4\pi G \rho_{m}\delta = 0.
\end{equation}
One can solve these two equations numerically. To fix the initial conditions, we assume that in  early times,the universe is matter dominated with negligible dark energy contribution which is typical for thawing model. For the matter dominated regime, $\delta \sim a$ and ${d\delta\over{da}} \sim 1$ fixes our initial conditions.  In figure 2, we plot the evolution of both  linear as well as nonlinear density contrast as a function of scale factor. We show the behaviors for scalar as well as tachyon field with linear and PNGB potential. To compare the results with $\Lambda$CDM, we also plot the corresponding behavior for $\Lambda$CDM case. As one can see from this plot, that although the linear matter density contrasts for different thawing models do not deviate much from that of the  $\Lambda$CDM model but the nonlinear density contrast has substantial deviation from $\Lambda$CDM model for both scalar as well as tachyon field. We should mention that although we show the results for linear and PNGB potentials, for other potentials, the behaviors are similar but with lesser deviations from $\Lambda$CDM model.
\vspace{5mm}

\section{The Spherical Collapse Model}
The spherical collapse model, which has a long history in cosmology, is a simple and a fundamental tool for understanding how a small spherical patch of over density forms a bound system via gravitational instability \citep{Gunn1972}.
As we consider flat, homogeneous and isotropic cosmologies, driven by non relativistic matter and dark energy with the equation of state, $p_\phi=w(a)\rho_\phi$ with $ p_\phi < 0$, the equations that describe our background universe are given by

\begin{figure}[t]
\includegraphics[width=80mm]{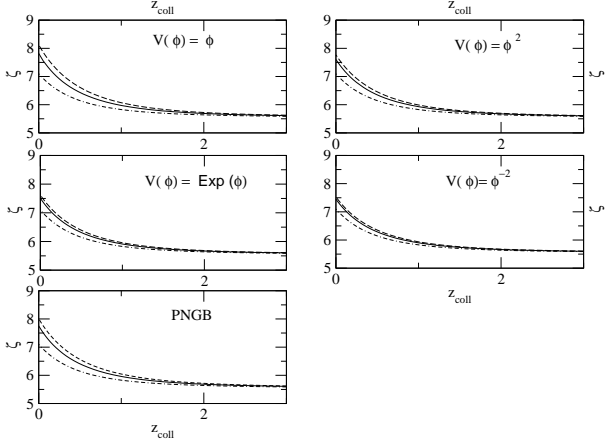}
\caption{Over density at turnaround vs. collapse redshift($z_{coll}$) for different scalar field
 models with different potentials  $V(\phi) = \phi, \phi^2, e^{\phi}, \phi^{-2}$ for  $\Omega_{m0} = 0.25$. In each figure, solid curve represents inhomogeneous case, dashed curve represents homogeneous case and dotdashed represents $\Lambda$CDM model. Scalar field PNGB model is also compared with $\Lambda$CDM model in the bottom left figure. }
\end{figure}

\begin{equation}
\left( {\dot{a}\over{a}}\right) ^2={8 \pi G\over{3}}({\rho_m}+{\rho_{\phi}})
\end{equation}and
\begin{equation}
{\ddot{a}\over{a}}=-{4\pi{G}\over{3}}[(3w(a)+1)\rho_{\phi}+{\rho_m}]
\end{equation}
 where $a(t)$ is the scale factor, $\rho_m=\rho_{m0}a^{-3}$ is the background matter density and $\rho_{\phi}=\rho_{\phi0}{f(a)}$ is the dark energy density, with 
  \begin{equation}
 {f(a)}= Exp[3\int_{a}^{1}\left( \frac{{1+w(u)}}{u}\right) {\rm d}u].
  \end{equation}
 The Friedmann equation can be written more simply as
 $H^{2}\equiv (\dot{a}/a)^{2}=H_{0}^2E(a)^2$ and $H_{0}$ is the Hubble constant with
 \begin{equation}
E(a)=\left[ \Omega_{\rm m0}a^{-3}+
\Omega_{\phi0}f(a)\right]^{1/2} \;\;\;,
\end{equation}
while $\Omega_{\rm m0}=  (8 \pi G\rho_{\rm m0})/3H_{0}^{2}$ 
 is the matter density parameter and
$\Omega_{\phi0}= (8 \pi G\rho_{\phi0})/3H_{0}^{2}$ 
 is the corresponding dark energy parameter at the present epoch with
$\Omega_{\rm m0}+\Omega_{\phi0}=1$.
The $\Omega_{m}(a)$ and $\Omega_{\phi}(a)$ evolve 
with the scale factor as
\begin{equation}
\Omega_{\rm m}(a)=\frac{\Omega_{\rm m0} 
a^{-3}}{E^{2}(a)} \;\;\;{\rm and} \;\;\; 
\Omega_{\phi}(a)=\frac{\Omega_{\phi0} f(a)}{E^{2}(a)} .
\end{equation}
Once we know the solution for $\Omega_{\phi}(a)$and $\gamma(a)$ by solving either
(5)-(7) or (13)-(15), we can easily find the behavior of the normalized
Hubble parameter in terms of $\Omega_{\phi}$ as
\begin{equation}
E^2(a)=\frac{H^2(a)}{H_0^2}=\frac{1-\Omega_{\phi 0}}{1-\Omega_{\phi}} a^{-3},
\label{E}
\end{equation}
 and the equation of state of dark energy $w(a)=\gamma(a)-1$.

\begin{figure}[t]
\includegraphics[width=80mm]{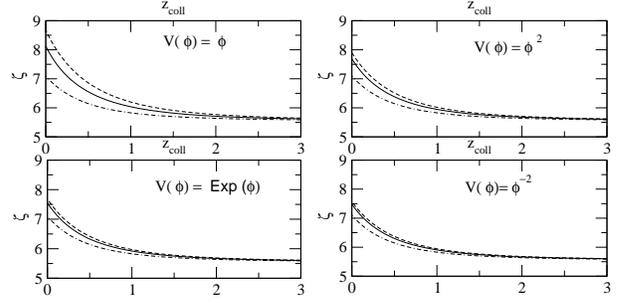}
\caption{Over density at turnaround vs. collapse redshift ($z_{coll}$) for different tachyon field models with the same potentials as in fig.3 for  $\Omega_{m0} = 0.25$. The lines correspond to the same as in fig.3.}
\end{figure}

The evolution of a spherical over dense patch of radius R(t) in the presence of dark energy is given by the Raychaudhuri equation:
\begin {equation} 
 {\ddot{R}\over{R}}=-4 \pi G\left[ \left(w(R)+{1\over{3}}\right) {\rho_{\phi c}}+{1\over{3}}{\rho_{m c}}\right] 
 \end{equation} 
where $\rho_{\phi c}$ and $\rho_{m c}$ are the  dark energy density and the matter density inside the spherical cluster respectively. $w(R)$ represents the equation of state of the dark energy inside the cluster. With some initial over density, the spherical over dense region will expand until it reaches the maximum radius (turn around point, $\dot{R}= 0$) and then begins to collapse. After that the sphere virializes forming a bound system. With different initial density, it happens locally at different regions of the universe. Now, after performing the following transformations: 
\begin{equation}
  x=\frac{a}{a_t}\;{\rm and}\;y=\frac{R}{R_t} 
\end{equation}
the equation of background evolution and that of the spherical perturbation become:
\begin{equation}
  \dot{x}^2={H_t}^2\Omega_{m,t}[ \Omega_m(x)x]^{-1}
 \end{equation}
 and
 \begin{equation}
{\ddot y}=-\frac{H_{\rm t}^{2}\Omega_{\rm m,t}}{2}
\left[ \frac{\zeta}{y^{2}}+\nu y I(x,y)\right] \;\;\;
\end{equation} 
where
\begin{equation}
I(x,y)=\left\{ \begin{array}{cc}
       \left[1+3w(R(y))\right]\frac{f(R(y))}{f(a_t)} &
       \mbox{Clustered DE}\\
       \left[1+3w(x)\right]f(x) & \mbox{Homogeneous DE}
       \end{array}
        \right.
\end{equation}
with
\begin{equation}
\nu=\frac{\rho_{\rm \phi, t}}{\rho_{\rm m, t}}=\frac{1-\Omega_{\rm m,
    t}}{\Omega_{\rm m, t}} \;\;\;.
\end{equation}
 
Subscript ``t'' denotes the turn around time. In order to solve the above set of equations, we have used the fact that the mass of the forming cluster is conserved:
$\rho_{\rm mc}R^3=\rho_{\rm mc,t}R_{\rm t}^3$ and define the relation:
\begin{equation} 
\rho_{\rm mc}=\rho_{\rm mc, t} \left(\frac{R}{R_{\rm t}}\right)^{-3}=
\frac{\zeta \rho_{\rm m, t}}{y^{3}}
\end{equation}
Here $\zeta$ is the matter density contrast at turnaround which is defined as
 \begin{equation}
\zeta \equiv{\rho_{mc,t}\over{\rho_{m,t}}}=\left( R_t\over{a_t}\right) ^{-3}.
 \end{equation}
 The function $R(y)$ is given by $R(y)=R_{\rm t}y=\zeta^{-1/3} a_{\rm t}y$ and $\Omega_{m} (x)$ by
\begin{equation}
\Omega_{\rm m}(x)=\frac{1}{1+\nu x^{3} f(x)} .
\end{equation}
Typically the scalar fields we consider here, have extremely small masses (of the order of present day Hubble scale in natural units). This is necessary to have nearly flat potentials for the scalar fields at present day. Due to this, the scale of fluctuations for these scalar fields are extremely  large, making it a smoothly distributed field  within the horizon scale. So it is safe to assume the dark energy to be homogeneous. But it is still interesting to consider the case where the dark energy clusters along with the dark matter and avoid the energy non-conservation problem examined in \citep{Maor2005}.  Hence in our subsequent calculations, we assume both the cases mentioned in equation (28).

\begin{figure}[t]
\includegraphics[width=80mm]{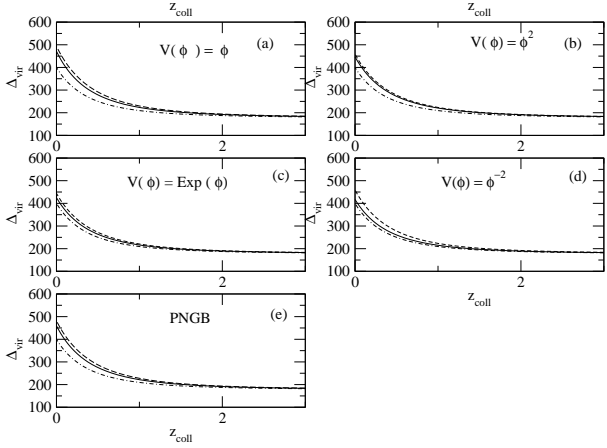}
\caption{Plot of the non-linear density contrast at virialization as the function of collapse redshift ($Z_{coll}$) for different Scalar 
field models with $V(\phi) = \phi, \phi^2, e^{\phi}, \phi^{-2}$ for  $\Omega_{m0} = 0.25$. In each figure, solid curve represents inhomogeneous case, dashed curve represents homogeneous case and dotdashed represents $\Lambda$CDM model. Scalar field PNGB model is also compared with $\Lambda$CDM model in the bottom last figure.}
\end{figure}

When dark energy is clustered together with the matter, one can solve analytically the system of equations (26) and (27)as the function $I(x,y)$ in eq.(28) depends only on $y$. So we easily calculate the density contrast $\zeta$ at the turn around epoch by integrating the equations (26) and (27) and applying the boundary conditions $(dy/dx)_{x=1} = 0$ and $y_{x=0} = 0$. The corresponding integral equation which
governs the behavior of $\zeta$, for a flat
cosmological model is given below:
\begin{eqnarray}
\int_{0}^{1} \left[\frac{y}{\zeta+\nu yP(y)-(\zeta+P(1)\nu)y}\right]^{1/2}{\rm d}y =\nonumber \\
\int_{0}^{1} \left[x\Omega_{\rm m}(x)\right]^{1/2}{\rm d}x
\end{eqnarray}
 
In the case of homogeneous dark energy, one can not follow this analytical procedure and has to solve the system of equations (26) and (27) numerically with the boundary conditions $(dy/dx) _{x=1} = 0$ and $y_{x=1} = 1$. 

In figure 3 and 4, we show the matter density contrast at turnaround,  $\zeta$ as a function of collapsed redshift for scalar and tachyon models with different potentials. To compare our result with $\Lambda$CDM model, we also show the corresponding $\zeta$ for $\Lambda$CDM model. At higher $z_{coll}$, overdensities at turnaround for different models tend toward the fiducial value of $5.6$ for Einstein-de Sitter universe. For structures which are collapsing around present time i.e $z_{coll} \sim 0$, the over densities at turnaround  for different scalar and tachyon models are higher than the corresponding $\Lambda$CDM value. Also the values for the homogeneous case are slightly  higher than the inhomogeneous case and the difference is more prominent for  tachyon model with linear potential. For other cases the differences are extremely small. 

  Just to mention, we show all the plots assuming $\Omega_{m0} = 0.25$. For other values of $\Omega_{m0}$, the behaviors are similar with lesser deviation from $\Lambda$CDM for higher $\Omega_{m0}$ and vice versa.

\begin{figure}[t]
\includegraphics[width=80mm]{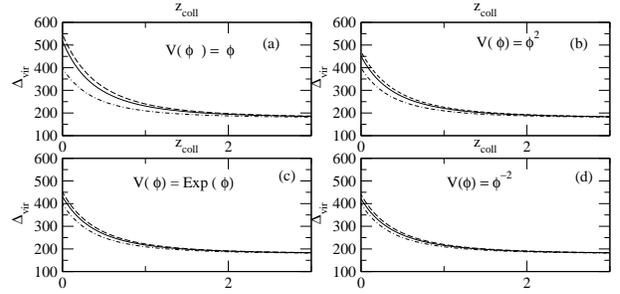}
\caption{ The plot of the non-linear density contrast at virialization as the function of  collapse redshift($z_{coll}$) for different tachyon field models for $\Omega_{m0} = 0.25$ with the same potentials as in fig.5. The lines correspond to the same as in fig.5.}
\end{figure}

\subsection{VIRIALIZATION IN THE SPHERICAL MODEL}
 
\begin{table*}
\centering
\begin{tabular}{|c|c|c|c|c|c|}

\hline
Model & Potential & Case & a & b & n  \\
\hline 
   &  $\phi $  & inhom &$ 0.88\pm 0.02$ &  $0.34 \pm 0.03$ &$ 1.34 \pm 0.07$\\ 
        &          & hom  & $0.88 \pm 0.02$ &  $0.42 \pm 0.02$ & $1.25 \pm 0.05$\\
\cline {2-6}
 & $\phi^{2} $& inhom &$ 0.855 \pm 0.004$ &$ 0.383 \pm 0.006$ &$ 1.10 \pm 0.02$\\
Tachyon  &          & hom & $0.853\pm 0.004$ & $0.432 \pm 0.005$ & $1.06 \pm 0.01$\\
        \cline {2-6}
 & $\phi^{-2}$ & inhom &$ 0.8472 \pm 0.0003 $ & $0.3867 \pm 0.0005$ & $1.001\pm 0.001$\\
        &          & hom &$ 0.8462 \pm 0.0001$ & $0.4254 \pm 0.0002$ &$ 0.9824 \pm 0.0005$\\
        \cline{2-6}
 &$ e^{\phi}$ & inhom & $0.850 \pm 0.001$ &$ 0.387 \pm 0.002$ &$ 1.032 \pm 0.005$ \\
       &           & hom &$ 0.848 \pm 0.001 $& $0.429 \pm 0.001$ & $1.007 \pm 0.003$\\
\hline 
 & $ \phi $  & inhom & $0.862 \pm 0.008$ & $0.37 \pm 0.01$ & $1.16 \pm 0.03 $\\   
       &          & hom & $ 0.859 \pm 0.007$ & $0.43 \pm 0.01$ & $1.11 \pm 0.02$\\
\cline{2-6}      
 & $ \phi^{2}$ & inhom & $ 0.852 \pm 0.003$ & $ 0.333 \pm 0.004$ & $1.08 \pm 0.01$ \\
       &          & hom & $0.850 \pm 0.002$ & $0.431 \pm 0.003$ & $1.024 \pm 0.005$ \\
       \cline{2-6}
Scalar&  $\phi^{-2}$ & inhom & $ 0.8460\pm 0.0002 $ & $0.3865 \pm 0.0003 $& $0.9867 \pm 0.0007$\\
       &          & hom & $ 0.8453 \pm 0.0006$ & $0.4241 \pm 0.0008 $ & $0.972 \pm 0.002 $\\
       \cline{2-6}
& $ e^{\phi} $ & inhom & $0.8480 \pm 0.0007$ & $0.387 \pm 0.001$ & $1.011 \pm 0.002$ \\
      &           & hom & $0.8469 \pm 0.0002$ & $ 0.4275 \pm 0.0003$ & $ 0.9905 \pm 0.0006$\\
      \cline{2-6}
 & PNGB & inhom & $0.858 \pm 0.006$ & $ 0.381 \pm 0.008$ & $ 1.12 \pm 0.02$\\
       &      & hom & $ 0.855 \pm 0.005$ & $ 0.434 \pm 0.007$ & $1.08 \pm 0.01$\\
\hline 
\end{tabular} 
\caption{ Values of the fitting parameters $a,b$ and $c$ by fitting equation (44) with the exact results given by equation (40) for scalar and tachyon models. We have also shown the errors for different parameters at $95\%$ confidence level} 
\end{table*}

In general the spherical collapse formalism leads to a point singularity as the final state of the system. But physically the objects go through a virialization process and stabilize to a finite size. Such process of virialization is not built in the spherical collpase formalism but we have to put it by hand in order to ensure virialization.

There are number of things that enter in this process of virialization as  we describe below.

First of all, the total energy has to be conserved always, hence the total energy at the time of turnaround should be equal to that at the time of virialization. At the turnaround the potential energy only contributes to the total energy of the system whereas the point of virialization is defined as where the virial theorem holds i.e the kinetic energy is related to the potential energy $T_{vir} = {1\over{2}}(r {\partial U\over\partial {r}})_{vir}$, where $T$ and $U$ are the kinetic and potential energy respectively. In our subsequent calculations we investigate two specific cases mentioned below:

\begin{itemize}
\item In the first case, we assume the dark energy is inhomogeneous and it virializes together with the matter. This assumption affects the way one calculates the turnaround point, as mentioned in equation (28) ( the case ``Clustered D.E''). On the other hand, the assumption that the dark energy takes part in the virialization process demands that the virial theorem should hold for the total kinetic and potential energy of the system (i.e matter + dark energy). Thus, applying the energy conservation together with this, we get\citep{Basilakos2007}

\begin{equation}
T_c = -\frac{1}{2}U_{Gc}+{U}_{\phi c}
\end{equation}
 
\begin{equation}
 {T}_{c,f}+{U}_{G c,f}+{U}_{\phi c,f} = {U}_{G c,t}+{U}_{\phi c,t}
\end{equation}
 where ${T}_c$ is the kinetic energy and ${U}_{Gc}=-3{G}M^2/5R$ is the potential energy for matter, $M$ being the matter mass contained in the spherical over density. $U_{\phi c} = -(1+3w)\rho_{\phi_{c}}{4{\pi}GM\over{10}}R^2$ is the potential energy associated with the dark energy inside the spherical over density. Here the subscripts ``f'' and ``t'' indicate the virialization and turn around time respectively and ``c'' for inside the cluster.
Using the above formulation, one can obtain a cubic equation which relates the ratio between the final (virial) $R_f$ and the turn-around radius $R_t$, defined as the collapse factor ${\lambda}=\frac{R_f}{R_t}$:
\begin{equation}
 2n_1{\lambda^3}-(2+n_2){\lambda}+ 1 = 0
\end{equation}where
\begin{equation}
 n_1=-(3w(a_f)+1){{\Omega_{\phi 0}}f(a_f)\over{\zeta{\Omega_{m0}}a_t^{-3}}}
\end{equation} and
\begin{equation}
 n_2=-(3w(a_t)+1){{\Omega_{\phi 0}}f(a_t)\over{\zeta{\Omega_{m0}}a_t^{-3}}}
\end{equation}with
$\rho_{\rm mc, t}$ being the matter density inside the sphere at the turn around time while 
$\rho_{\rm m, t}$ is the background matter density at the same epoch.

\item In the second case, we assume that the dark energy is homogeneous and it does not virialize inside the cluster. In this case, to get the turnaround point, we use eqn(28) (with the choice ``Homogeneous DE''). On the other hand, as the dark energy does not virialze, its only effect is to contribute to the potential energy of the system. So applying the energy conservation, we get\citep{Maor2005}
\begin{equation}
 \left[U_{\phi c}+ U_{Gc} + \frac{R}{2}\frac{\partial U_{Gc}}{\partial R}\right]_{f} = \left[U_{Gc}+U_{\phi c}\right]_{t}
\end{equation}
\end{itemize}


Assuming that at the collapse point, the system has virialised fully, the density contrast at virialization as a function of $z_{coll}$ and $\Omega_{m0}$ is given by 
\begin{equation}
\Delta_{\rm vir}=\frac{\rho_{\rm mc, f}}{\rho_{\rm m, f}}=
\frac{\zeta}{\lambda^{3}} 
\left(\frac{a_{\rm f}}{a_{\rm t}}\right)^{3},
\end{equation}
where the relation 
between  $a_{\rm f}$ and $a_{\rm t}$ can be estimated from the equation,
\begin{equation}
\frac{dt}{da}=\frac{1}{H(a) a} \;\;,
\end{equation}
together with the condition that  the time needed 
to collapse is twice the turn-around time, $t_{\rm f}=2t_{\rm t}$, i.e.
\begin{equation}
\int_{0}^{a_f}\frac{1}{H(a)a}{\rm d}a=2\int_{0}^{a_t}\frac{1}{H(a)a}{\rm d}a.
\end{equation}
In case of a $\Lambda$ cosmology we get an analytical solution:
 \begin{equation}
 \sinh^{-1}({a_f}^{3/2}\sqrt{\nu_0})=2\sinh^{-1}({a_t}^{3/2}\sqrt{\nu_0}),
 \end{equation}
 where $\nu_0=(1-\Omega_{m0})/\Omega_{m0}$. The ratio between the scale factors converges to the Einstein de Sitter value $(\frac{a_f}{a_t})=2^{2/3}$ at high redshifts.

In figure 5 and 6, we show the behavior of matter density contrast at virialization $\Delta_{vir}$ for different thawing models as a function of collapsed redshift $z_{coll}$. This parameter plays a crucial role as an observational tool which can be directly applied to Press-Schechter theory for comoving number density of virialized objects. The behavior is consistent with the fact that with $z_{coll}$, $\Delta_{vir}$ decreases and settles to the value $18 \pi^2$ for Einstein de-Sitter universe.

\begin{figure*}
\includegraphics[width=6.6in]{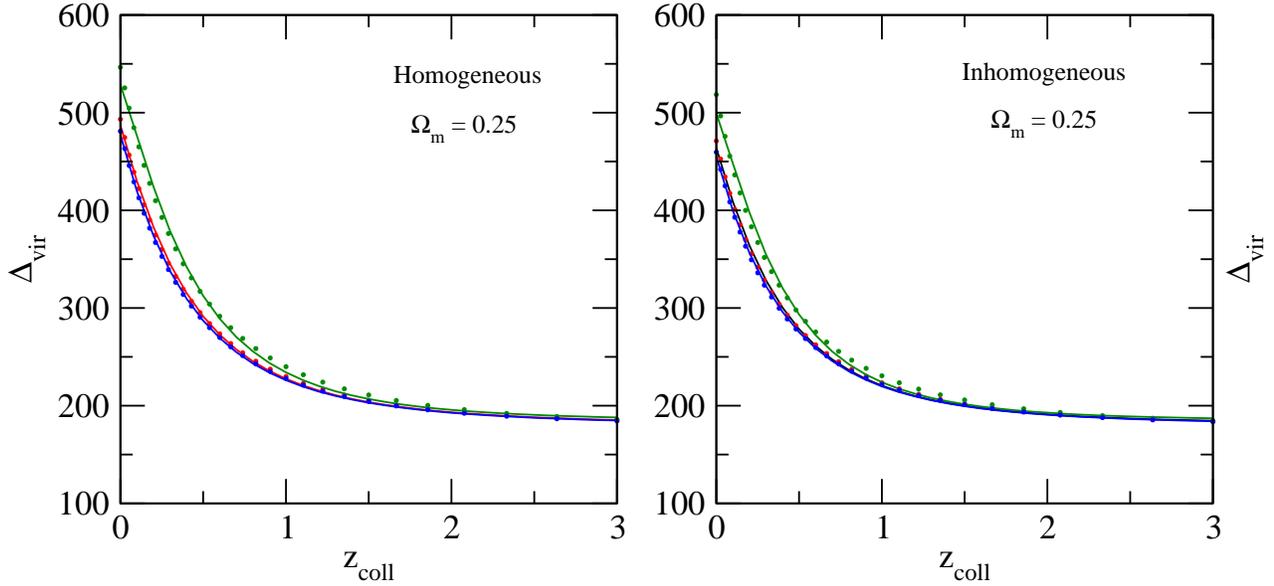}
\caption{Plot of the fitting function for the density contrast at virialization $\Delta_{vir}$ vs the collpase redshift ($z_{coll}$) for different models. From top to bottom represents tachyon with linear potential, scalar field with linear potential and scalar field with PNGB potential. For each case, the smooth line is for the fitting equation given by eqn (44) with the parameters mentioned in table 1, whereas the dots are generated by the exact numerical result for $\Delta_{vir}$ calculated for each model. }
\end{figure*}

Once virialized, the density contrast inside the cluster is slightly higher in the homogeneous case compared to the inhomogeneous case. This is consistent with the result at the turnaround where also similar things happen. This means that in the homogeneous case, where only matter virializes, the density inside the clustered objects are higher. This is in agreement with the result obtained by  Maor \citep{Maor2006} for dark energy with constant equation of state. There are also few interesting results. The difference in $\Delta_{vir}$ for a particular thawing model with the corresponding $\Lambda$CDM model is larger for scalar and tachyon field with linear potential as well as for scalar field with PNGB potential. Also tachyon field with linear potential has the largest deviation from $\Lambda$CDM.  This predicts that these models are easier to distinguish from $\Lambda$CDM model by observing abundances of bound objects. For scalar field model, the diference between homogeneous and inhomogeneous model is highest for the potential $V(\phi) = {1\over{\phi^2}}$. It is interesting to note that this potential also can act as a tracker or freezing model. For tachyon model, these two cases are hardly distinguishable for all of the potentials. 

Given the fact that the mass of the dark energy has to be extremely small, it is most likely that the field responsible for dark energy has to be homogeneous and should not cluster inside the virialized objects. If this is the scenario which is physically relevant, our results show that scalar as well as tachyon field dark energy models can be distinguished from the $\Lambda$CDM model by observing the abundances of bound objects since the virialized density contrast for homogeneous cases deviate sufficiently from the $\Lambda$CDM model.

We should mention, that the final results for $\Delta_{vir}$, depend on both the background evolution as well as the fact that dark energy may cluster. The fact that the $\Delta_{vir}$ is slightly smaller for inhomogeneous case than the homogeneous one for all potentials, shows that the inhomogeneous dark energy acts against the matter clustering. But as the model is quite nonlinear and involves rigorous numerical computations, it is very difficult to predict the percentage of effects coming from the background evolution as well as from dark energy clustering which may be an useful information  for actual N-body simulations.        

Next, we give a fitting formula for $\Delta_{vir}$ as a function of collapsed redshift $z_{coll}$ which is given by
\begin{equation}
\Delta_{vir}(z_{coll}) = 18\pi^2\left(a + b* \Theta(z_{coll})^n\right)
\end{equation}

where $\Theta(z_{coll}) = {1\over{\Omega_{m}(z_{coll})}} -1$. In Table 1, we show the values of the fitting parameters $a,b$ and $n$ for different models assuming $\Omega_{m0} = 0.25$. We also quote the the corresponding errors for these parameters at the $95\%$ confidence level.

In figure 7, we show accuracy of our fitting function for thawing models with linear as well as PNGB potential potential. For other potentials it also works equally well. 

\begin{figure*}
\includegraphics[width=6.6in]{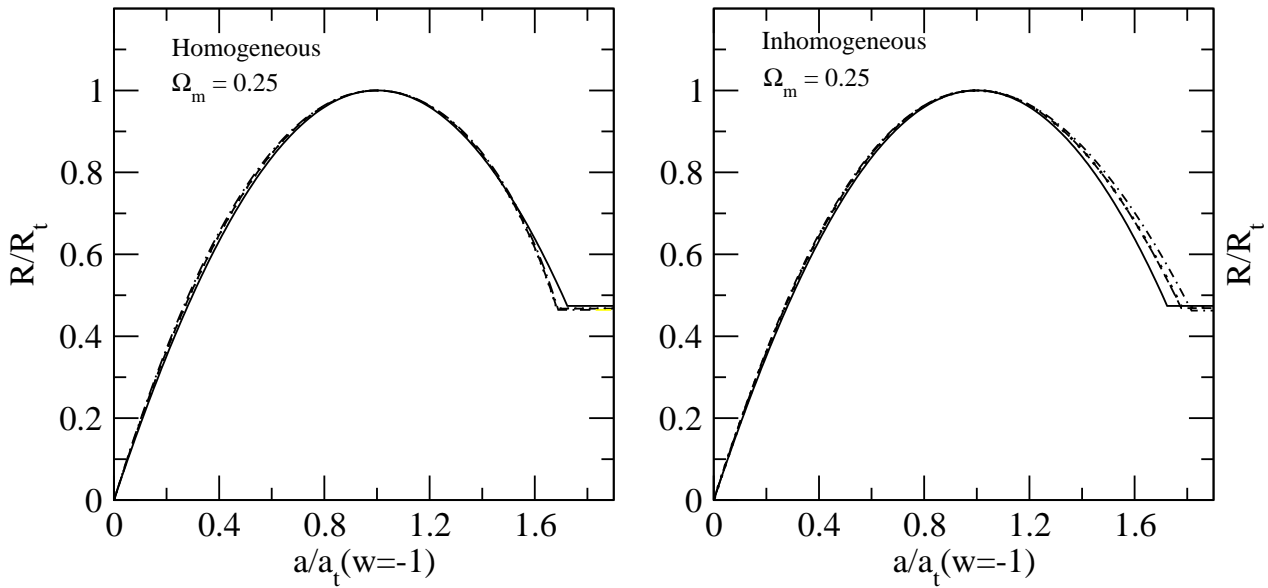}
\caption{Variation of the radius $R$ normalized to the turnaround radius $R_{t}$ w.r.t. the scale factor normalized to the turnaround scale factor for the $\Lambda$CDM model. Solid line is for $\Lambda$CDM whereas the dashed, dash-dotted and dotted are for scalar field woth linear potential, tachyon with linear potential and scalar field with PNGB potential.}
\end{figure*}

Knowing that models with linear potential (for both tachyon and ordinary scalar fields) as well as models with PNGB type potential deviate more from $\Lambda$CDM model in comparison to other potentials, we study the size of the collapsed objects for these potentials for homongeneous as well as for inhomogeneous cases. In figure 8, we show the evolution of the perturbation collapsing radii at the present time ($z_{coll} = 0$) for these potentials. The radius has been normalized to the turnaround radius $R_{t}$. It has been plotted with respect to the scale factor normalized to the turnaround scale factor for $\Lambda$CDM model.  The collapse to  singularity is avoided assuming that after the virialization, the size becomes constant. From the figure 8, for homogeneous case, virialization occurs earlier than $\Lambda$CDM model whereas for inhomogeneous case, it happens later. This means, homogeneous models are more suitable for producing older bound objcets than $\Lambda$CDM. This gives a slight edge to homogeneous dark energy models over the $\Lambda$CDM model.

Although, in figure 8, the virialized radii for different models are practically indistinguishable but still there are slight differences. First of all, the virialized radii for different scalar and tachyon models are smaller than the $\Lambda$CDM model, which in turn confirms that objects are more denser in these models than $\Lambda$CDM model. This is consistent with the results shown in figure 5 and 6. Secondly between individual models, virialised radius for tachyon model with linear potential is smaller than the other two models, e.g, scalar field with linear potential and scalar field with PNGB potential. This is also consistent with behaviour of virialized density contrast shown in figure 7. Also the virialized radii in inhomogeous models are slightly larger than the homogneous counterpart, making the homogeneous one denser than inhomogeneous one. This is also consistent with what we show in figure 5 and 6.

\section{Conclusion}

In this work, we study the evolution of spherical overdensities for a general class of thawing dark energy models. We consider both ordinary scalar fields as well as tachyon fields having noncanonical kinetic energy. We consider variety of potentials for the fields. In studying the evolution of spherical over densities, we consider the dark energy  to be homogeneous as well as inhomogeneous. For the homogeneous case, we assume only matter virializes inside the cluster whereas for the inhomogeneous case, we assume both matter and dark energy virializes inside the cluster.

Our main motivation is to see whether one can distinguish thawing dark energy models from $\Lambda$CDM model studying the evolution of spherical over density. Our results show that although all the models become indistinguishable for objects collapsing earlier, for objects collapsing around present time, some of the thawing models deviate significantly from the $\Lambda$CDM model. To be specific, models with linear potential both for standard scalar fields and tachyon, together with scalar field with PNGB type potential have significant deviation from $\Lambda$CDM model. We have shown that the size of the virialised object, the density contrast at turnaround as well as at virialization differ significantly for these models from $\Lambda$CDM model. This is consistent with the results obtained by Sen et al. \citep{soma2010} for thawing models considering the background evolution.  Also the deviations are enhanced in the homogeneous dark energy case where matter only virializes inside the cluster, thereby making it more probable to be distinguished from $\Lambda$CDM. Given the fact that the mass of the dark energy field has to be extremely small, it is safe to assume that the dark energy remains homogeneous for most of the relevant scales. Hence the evolution of spherical overdensities in large scale structure formation can be a useful tool to distinguigh model from $\Lambda$CDM.

 We have also derived the fitting formula for $\Delta_{vir}$ as function of collapsed redshift for different models and tabulated the values of the fitting parameters for different models assuming the dark energy to be homogeneous as well as inhomogeneous. Although we have not shown explicitly, if one varies $\Omega_{m0}$, deviations are higher for values smaller than  $\Omega_{m0} = 0.25$ and vice versa. This is because smaller the value of $\Omega_{m0}$, higher the contribution from dark energy at late times and hence one needs more matter inside the cluster for objects collapsing at late times.

An important conclusion one can draw from this study  is that thawing models with linear potential can have significant deviation from $\Lambda$CDM model. Now, scalar fields with linear potential have received little attention as a viable dark energy candidate \citep{Kratochvil2004}. But it certainly deserves more attention, especially for the homogeneous case.

Finallly, we want to stress that we study a toy model for nonlinear evolution of over density where we use simplified assumption of spherical over-densities. In order to compare the theoretical results with observational data, one has to give away this simplified assumption and has to perform full N-body simulations to calculate accurately the dark energy signatures in nonlinear structure formation. But even with this simplified assumption, we find some interesting features in  the nonlinear evolution of matter over-densities in the thawing class of dark energy models. This may be helpful for more detailed N-body simulations. 

\section{Acknowledgment}

The authors acknowledge the financial support provided by the University Grants Commission, Govt. Of India, through the major research project grant (Grant No:  33-28/2007(SR)).


\begin{thebibliography}{99}
\bibitem[\protect\citeauthoryear{Abrahamse}{2007}]{Abrahamse2007}Abrahamse~A., et al., 2007,~Phys.\ Rev.\ D, 77, 103503

\bibitem[\protect\citeauthoryear{Abramo \& Finelli}{2003}]{Abramo2003}Abramo L.~R.~W. \& Finelli F., 2003,
Phys.\ Lett.\ B, 575, 165 

\bibitem[\protect\citeauthoryear{Acquaviva \& Baccigalupi}{2006}]
{Acquaviva 2006}Acquaviva~V. ~\& Baccigalupi~C., 2006, Phys. Rev. D, 74, 103510 

\bibitem[\protect\citeauthoryear{Aguirregabiria \& Lazkoz}{2004}]
{Aguirregabiria2004}Aguirregabiria J.~M. ~\& Lazkoz~R., 2004,~Phys.\ Rev.\ D, 69, 123502

\bibitem[\protect\citeauthoryear{Alam}{2003}]{Alam2003}Alam~U., Sahni~V. ~\& Staronbinsky~A.A.,~2003,~JCAP, 0304, 002



\bibitem[\protect\citeauthoryear{Ali}{2009}]{Amna2009}Ali~A.,~Sami~M. ~\& Sen A.A., 2009, Phys.\ Rev.\ D, 79, 123501

\bibitem[\protect\citeauthoryear{Allen}{2004}]
{Allen2004}Allen~S.~W.,~ et.al., 2004, MNRAS., 353, 457 

\bibitem[\protect\citeauthoryear{Avelino}{2008}]{Avelino2008}Avelino~P.~P.,~Beca L.~M.~G. ~\& Martins~C.~J.~A.~P., 2008, Phys.\ Rev.\  D, 77, 101302 

\bibitem[\protect\citeauthoryear{Baccigalupi \& Acquaviva}{2006}]
{Baccigalupi2006}Baccigalupi~C. ~\& Acquaviva~V., 2006, (astro-ph/0606069)

\bibitem[\protect\citeauthoryear{Bagla }{2003}]
{Bagla2003}Bagla J.~S.,~Jassal H.~K. \& Padmanabhan. T., 2003,
Phys.\ Rev.\ D, 67, 063504 

\bibitem[\protect\citeauthoryear{Baldi}{2008}]
{Baldi2008}Baldi~M., et al., 2008, (arXiv:0812.3901) 

\bibitem[\protect\citeauthoryear{Bartelmann}{2006}]{Bartelmann2006}Bartelmann M., Doran M.  \& Wetterich C., 2006, A\& A, 454, 27

\bibitem[\protect\citeauthoryear{Bartolo}{2004}]{Bartolo2004}Bartolo~N., Corasaniti~P. S., Liddle~A. R. ~\& Malquarti~M., 2004, Phys.\ Rev.\ D, 70, 043532 


\bibitem[\protect\citeauthoryear{Basilakos}{2003}]{Basilakos2003}Basilakos S., 2003, ApJ, 590, 636


\bibitem[\protect\citeauthoryear{Basilakos \&  Voglis}{2007}]
{Basilakos2007}Basilakos~S. ~\& Voglis N., 2007, MNRAS, 374, 269, 
 

\bibitem[\protect\citeauthoryear{Bergshoeff}{2000}]{Bergshoeff2000}Bergshoeff~E.~A., et al., 2000, JHEP, 0005, 009 

\bibitem[\protect\citeauthoryear{Bernstein \& Jain}{2004}]
{Bernstein2004}Bernstein~G.~M. ~\& Jain B., 2004, ApJ., 600, 17 

\bibitem[\protect\citeauthoryear{Bharadwaj}{2009}]
{Bharadwaj2009}Bharadwaj~S.~, ~Sethi~S.~K. \& Saini~T.~D., 2009, Phys. Rev. D, 79, 083538  

\bibitem[\protect\citeauthoryear{Caldwell}{1998}]{Caldwell1998}Caldwell~R. R., Dave~R. ~\& Steinhardt~P. J., 1998, Phys. Rev. Lett., 80, 1582

\bibitem[\protect\citeauthoryear{Caldwell}{2009}]{Caldwell2009}Caldwell R. R. \& Kamionkowski M., 2009 (arXiv:0903.0866)

\bibitem[\protect\citeauthoryear{Caldwell \& Linder}{2005}]{Caldwell2005}Caldwell~R. R. ~\& Linder~E.~V., 2005, Phys. Rev. Lett., 95, 141301

\bibitem[\protect\citeauthoryear{Copeland}{Copeland}{2005}]{Copeland2005}Copeland~E. J., Garousi~M. R., Sami M., Tsujikawa S., 2005, Phys. Rev. D, 71, 043003

 
\bibitem[\protect\citeauthoryear{Copeland}{2006}]{Copeland2006} Copeland E.J., Sami~M. \& Tsujikawa ~S., 2006, Int. J. Mod. Phys. D, 15, 1753 


\bibitem[\protect\citeauthoryear{Corasaniti}{2005}]{Corasaniti2005}Corasaniti~P~.S., Giannantonio~T. ~\& Melchiorri A., 2005, Phys.Rev. D, 71, 123521

\bibitem[\protect\citeauthoryear{Courtin}{2010}]
{Couetin2010}Courtin~J., et al., 2010,~(arXiv:1001.3425)

\bibitem[\protect\citeauthoryear{Davis}{2007}]{Davis2007}Davis T.~M., et al., 2007, ApJ, 666, 716

\bibitem[\protect\citeauthoryear{Eisenstein}{1998}]
{Eisenstein1998}Eisenstein~D.~J., et al., 1998, ApJ., 494, L1

\bibitem[\protect\citeauthoryear{Felder}{2002}]{Felder2002}Felder~G. N., Kofman L. \& Starobinsky A., 2002, JHEP, 0209, 026

\bibitem[\protect\citeauthoryear{Frieman}{1995}]{Frieman1995}Frieman~A., et al., 1995, Phys. Rev. Lett., 75, 2077

\bibitem[\protect\citeauthoryear{Frieman}{2008}]{Frieman2008}Frieman J., Turner M. \& Huterer D., 2008, Ann. Rev. Astron. Astrophys., 46, 385 

\bibitem[\protect\citeauthoryear{Garousi}{2000}]{Garousi2000}Garousi~M.~R., 2000, Nucl. Phys. B, 584, 284 

\bibitem[\protect\citeauthoryear{Giannantonio}{2008}]{Giannantonio2008}Giannantonio~T.,~ et al., 2008, Phys. Rev. D, 77, 123520 
 
\bibitem[\protect\citeauthoryear{Giovi}{2005}]
{Giovi2005}Giovi~F., et al., 2005, Phys. Rev. D, 71, 103009

\bibitem[\protect\citeauthoryear{Gordon \& Hu}{2004}]
{Gordon2004}Gordon~C. ~\& Hu. W., 2004, Phys.\ Rev.\ D, 70, 083003

\bibitem[\protect\citeauthoryear{Gordon \& Wands}{2005}] 
{Gordon2005}Gordon~C. ~\& Wands~D., 2005, Phys.\ Rev.\ D, 71, 123505 

\bibitem[\protect\citeauthoryear{Gunn \& Gott}{1972}]{Gunn1972}Gunn J. E. \& Gott J. R., 1972, ApJ, 
176, 1

\bibitem[\protect\citeauthoryear{Haiman~}{2000}]
{Haiman2000}Haiman~Z., et al., 2000, Astrophys. J., 553, 545 

\bibitem[\protect\citeauthoryear{Horellou \& Berge~}{2005}]{Horellou2005}Horellou C. \& Berge J., 2005, MNRAS, 360, 1393

\bibitem[\protect\citeauthoryear{Hu~}{2005}]{Hu2005}Hu~W., 2005, Phys. Rev. D, 7
1, 047301

\bibitem[\protect\citeauthoryear{Jain \& Taylor~}{2003}]
{Jain2003}Jain~B.  \& Taylor~A., 2003, Phys. Rev. Lett., 91, 141302 

\bibitem[\protect\citeauthoryear{Jassal~}{2009}]
{Jassal2009}Jassal~H.~K., 2009, Phys. Rev. D, 79, 127301 

\bibitem[\protect\citeauthoryear{Jassal~}{2010}]
{Jassal2010}Jassal~H.~K., 2010, Phys. Rev. D, 81, 083513

\bibitem[\protect\citeauthoryear{Kluson~}{2000}]{Kluson2000}Kluson~J., 2000, Phys. Rev. D, 62, 126003

\bibitem[\protect\citeauthoryear{Knop~}{2003}]{Knop2003}Knop R. A., et al., 2003, ApJ, 598, 102

\bibitem[\protect\citeauthoryear{Kratochvil~}{2004}]{Kratochvil2004}Kratochvil~J., et al., 2004, JCAP, 0407, 001 

\bibitem[\protect\citeauthoryear{Kutasov \&  Niarchos~}{2003}]{Kutasov2003}Kutasov~ D. ~\& Niarchos~V., 2003, Nucl. Phys. B, 666, 56 

\bibitem[\protect\citeauthoryear{Lahav}{1991}]{Lahav1991}Lahav O., Lilje P. B., Primack J. R. \& Rees M. J., 1991, MNRAS, 251, 128

\bibitem[\protect\citeauthoryear{Liddle \& Scherrer}{1999}] 
{Liddle1999}Liddle~A. R. ~\& Scherrer~R. J,~1999 Phys. Rev. D, 59, 023509 

\bibitem[\protect\citeauthoryear{Linder}{2008}]{Linder2008}Linder E.~V., 2008, Gen. Rel. Grav., 40, 329

\bibitem[\protect\citeauthoryear{Lokas \& Hoffman}{2001}]
{Lokas2001}Lokas E. L. ~\& Hoffman~Y., 2001, (astro-ph/018283) 

\bibitem[\protect\citeauthoryear{Maccio}{2004}]
{Maccio2004}Maccio~A.V., et al., 2004, Phys. Rev. D, 69, 123516 

\bibitem[\protect\citeauthoryear{Macorra}{2003}]
{Macorra2003}Macorra~ A.~de la., 2003, Phys. Rev. D, 67, 103511

\bibitem[\protect\citeauthoryear{ Manera \&  Mota}{2006}]{Manera2006}Manera~M. \& Mota~ D. F., 2006, MNRAS, 371, 1373

\bibitem[\protect\citeauthoryear{Maor}{2006}]{Maor2006}Maor I, 2006, (astro-ph/0602441).

\bibitem[\protect\citeauthoryear{Maor \&  Lahav}{2005}]{Maor2005}Maor~I. \& Lahav~O.~, 2005, JCAP, 7, 3

\bibitem[\protect\citeauthoryear{Mota \& Van de Bruck}{2004}]{Mota2004}Mota D. F. \& van de Bruck C., 2004 A \& A, 421, 71

\bibitem[\protect\citeauthoryear{Nunes \& Mota}{2006}]{Nunes2006}Nunes N. J. \& Mota D. F., 2006, MNRAS, 368, 751

\bibitem[\protect\citeauthoryear{Padmanabhan}{2003}]{Padmanabhan2003}Padmanabhan~T., 2003, Phys. Rep., 
380, 235 

\bibitem[\protect\citeauthoryear{Pogosian}{2005}]
{Pogosian2005}Pogosian~L., et al., 2005, Phys. Rev. D, 72, 103519

\bibitem[\protect\citeauthoryear{Press \& Schechter}{1974}]{Press1974}Press~W. H. \& Schechter~P., 1974, ApJ, 187, 425

\bibitem[\protect\citeauthoryear{Ratra \& Peebles}{1988}]
{Ratra1988}Ratra B. \& Peebles P. J. E., 1988, Phys. Rev. D, 37, 3406

\bibitem[\protect\citeauthoryear{Riess}{2004}]{Riess2004}Riess A. G., et al., 2004, ApJ, 607, 665 

\bibitem[\protect\citeauthoryear{Sahni \& Starobinsky }{2000}]{Sahni2000}Sahni~V. \& Starobinsky A. A., 2000, Int. J. Mod. Phys. D, 9, 373

\bibitem[\protect\citeauthoryear{Sami}{2004}]{Sami2004}Sami~M., Savchenko N. \& Toporensky A., 2004, Phys. Rev. D, 70, 123528

\bibitem[\protect\citeauthoryear{Sami}{2009}]{Sami2009}Sami~M., 2009, (arXiv:0904.3445)

\bibitem[\protect\citeauthoryear{Schaefer \&  Koyama}{2004}]
{Schaefer2008}Schaefer~B. M. \&  Koyama~K., 2008, MNRAS, 385, 411 

\bibitem[\protect\citeauthoryear{Scherrer \& Sen}{2008a}]
{Scherrer2008a}Scherrer~R. J. \& Sen A. A.,~2008a, Phys. Rev. D, 77, 083515

\bibitem[\protect\citeauthoryear{Scherrer \& Sen}{2008b}]
{Scherrer2008b}Scherrer~R. J. \& Sen~A. A.,~2008b, Phys. Rev. D, 78, 067303

\bibitem[\protect\citeauthoryear{Sen}{2002a}]{Sen2002a}Sen~A., 2002a, JHEP , 0204, 048 

\bibitem[\protect\citeauthoryear{Sen}{2002b}]{Sen2002b}Sen~A., 2002b, Mod. Phys. Lett. A, 17, 1797

\bibitem[\protect\citeauthoryear{Sen}{2010}]{soma2010}Sen S.,~Sen A. A. \& Sami~M., 2010, Phys. Lett. B, 686, 1

\bibitem[\protect\citeauthoryear{Seo \& Eisenstein}{2003}]{Seo2003}Seo~H.~J. ~\& Eisenstein~D.~J., 2003, ApJ,  598, 720 

\bibitem[\protect\citeauthoryear{Shaw \& Mota}{2008}]
{Shaw2008}Shaw~D. J. \& Mota~D. F., ApJS., 2008, 174, 277

\bibitem[\protect\citeauthoryear{Silvestri}{2009}]{Silvestri2009}Silvestri~A. \& Trodden M., 2009, Rep. Prog. Phys., 72, 096901

\bibitem[\protect\citeauthoryear{Steinhardt}{1999}]
{Steinhardt1999}Steinhardt~P. J., Wang~L. \&  Zlatev~I., 1999, Phys. Rev. D, 59, 123504

\bibitem[\protect\citeauthoryear{Tsujikawa \& Sami}{2004}]{Tsujikawa2004}Tsujikawa S. \& Sami M., 2004, Phys. Lett. B, 603, 113

\bibitem[\protect\citeauthoryear{Unnikrishnan}{2008a}]
{Unnikrishnan2008a}Unnikrishnan~S., 2008a, Phys. Rev. D, 78, 063007 

\bibitem[\protect\citeauthoryear{Unnikrishnan}{2008b}]
{Unnikrishnan2008b}Unnikrishnan~S., Jassal~ H.~K. \& Seshadri~T.~R., 2008b, Phys. Rev. D, 78, 123504

\bibitem[\protect\citeauthoryear{Wang \& Steinhardt}{1998}]{Wang1998}Wang L. \& Steinhardt P. J., 1998, ApJ,
508, 483

\bibitem[\protect\citeauthoryear{Wang}{2006}]{Wang2006}Wang P., ApJ, 2006, 640, 18

\bibitem[\protect\citeauthoryear{Weinberg \& Kamionkowski}{2003}]{Weinberg2003}Weinberg N. N. \& Kamionkowski M., 2003, ApJ, 341, 251

\bibitem[\protect\citeauthoryear{Wood-Vasey}{2007}]{Wood-Vasey2007}Wood-Vasey~W.~M., et al., 2007, ApJ, 666, 694 

\bibitem[\protect\citeauthoryear{Zeng}{2005}]{Zeng2005}Zeng D. \& Gao Y., 2005, (astro-ph/0505164)

\bibitem[\protect\citeauthoryear{Zimdahl}{2005}]
{Zimdahl2005}Zimdahl~W. \& Fabris J.~C., 2005, Class.\ Quant.\ Grav., 22, 4311




\end{thebibliography}
\end{document}